\documentclass{ifacconf}
\usepackage{caption}
\usepackage{graphicx}      % include this line if your document contains figures
\usepackage{natbib}        % required for bibliography
\usepackage{float} 
\usepackage{amsmath} 
\usepackage{booktabs}
\usepackage{xcolor}
\begin{document}
\begin{frontmatter}
%\title{Enhanced hydrodynamic modeling of offshore wind turbines using Morison’s equations with a frequency-dependent hydrodynamic coefficients} 
\vspace{-20pt}
\makebox[0pt][c]{\raisebox{1pt}{%
  \parbox{\textwidth}{\centering
    \small © 2025. This work has been accepted to IFAC for publication under a Creative Commons Licence CC-BY-NC-ND and will be presented at the Modeling, Estimation and Control Conference (MECC 2025) in Pittsburgh, Pennsylvania, USA.}
}}

\vspace{-15pt}

\title{Enhanced Hydrodynamic Modeling of Offshore Wind Turbines using Morison’s Equation with Frequency-Dependent Coefficients} 

\thanks[footnoteinfo]{ This work was supported by the U.S. Department of Energy
Advanced Research Projects Agency for Energy (ARPA-E) under
the grant DE-AR0001187.}

\author[First]{Md Sakif} 
\author[First]{Doyal Sarker} 
\author[First]{Kazi Mohsin} 
\author[First]{Tri Ngo} 
\author[First]{Tuhin Das} 

\address[First]{Department of Mechanical and Aerospace Engineering, University of
Central Florida, FL 32816, USA (e-mail: mdrafidulhaque.sakif@ucf.edu, doyal.kumar.sarker@ucf.edu,
 kazi.ishtiak.mohsin@ucf.edu,
 tri.ngo@ucf.edu, tuhin.das@ucf.edu)}

\begin{abstract}                % Abstract of 50--100 words

This paper presents a novel approach for implementing frequency-dependent hydrodynamic coefficients in Morison’s equation, which is widely used in hydrodynamics modeling. Accurate hydrodynamic predictions using Morison’s equation necessitate the incorporation of frequency-dependent drag coefficients due to their variation with wave frequency. To address this, the proposed method segments the frequency domain into different regions, such as low-frequency (resonance) and high-frequency (wave) regions. Instead of using a constant drag coefficient across the entire spectrum, different drag coefficients are assigned to these regions. To implement this, a fifth-order low-pass Butterworth velocity filter is applied for the resonance zone, while a first-order high-pass Butterworth velocity filter is applied for the wave-dominated zone. The approach is validated using the INO WINDMOOR 12MW semisubmersible offshore wind turbine, comparing the simulation results against the experimental data. By incorporating frequency-dependent drag coefficients, the model shows improved agreement with experimental surge motion data across both frequency regions, demonstrating the effectiveness of the proposed method.
\end{abstract}


\end{frontmatter}
%===============================================================================

\section{Introduction}

Offshore wind energy has emerged as a key area of focus in renewable energy research and development. In the past four years, the installed capacity of offshore wind turbines has seen significant growth, increasing from 28.29 GW to 72.66 GW \textbf{\cite{jung2024future}}, making them an increasingly significant contributor to global energy production. Among various technologies, floating offshore wind turbines (FOWTs) are especially promising for harnessing stronger and more consistent wind resources available in deep waters. As the demand for FOWTs continues to rise, optimizing their design becomes more essential since these large structures have relied on massive floating platforms for stability. To move beyond the limitations of massive floating platforms, innovative approaches such as Control Co-Design (CCD), \textbf{\cite{garcia2019control}}, are gaining attention. CCD minimizes excessive mass by integrating advanced control systems, optimizing both the controller and physical dynamics for a more efficient and cost-effective FOWT design. By incorporating all relevant engineering disciplines from the outset, CCD emphasizes feedback control and dynamic interactions as the key drivers of design. Developing next-generation computational tools to support CCD in FOWT design is essential.
OpenFAST, developed by NREL, \textbf{\cite{jonkman2022openfast}}, is a multi-physics engineering tool for simulating the coupled dynamic response of wind turbines. It integrates aero-hydro-elasto dynamics, and control and electrical system (servo) dynamics models, enabling fully coupled nonlinear aero-servo-elastic simulations in the time domain. Another simulation framework, the Control-oriented, Reconfigurable, and Acausal Floating Turbine Simulator (CRAFTS), has been developed to support the CCD of FOWT systems. It features a library-based, modular, and hierarchical architecture, allowing the integration of coupled-dynamics modules such as aerodynamics, structural dynamics, hydrodynamics, mooring, and control systems. Details of these modules in CRAFTS are discussed in \textbf{\cite{odeh2023development}}, \textbf{\cite{sarker2024modeling}}, \textbf{\cite{mohsin2025dynamically}}. 

\begin{figure*}[h]
\begin{center}
\includegraphics[width=18cm, trim={8cm 10cm 10cm 5cm},clip]{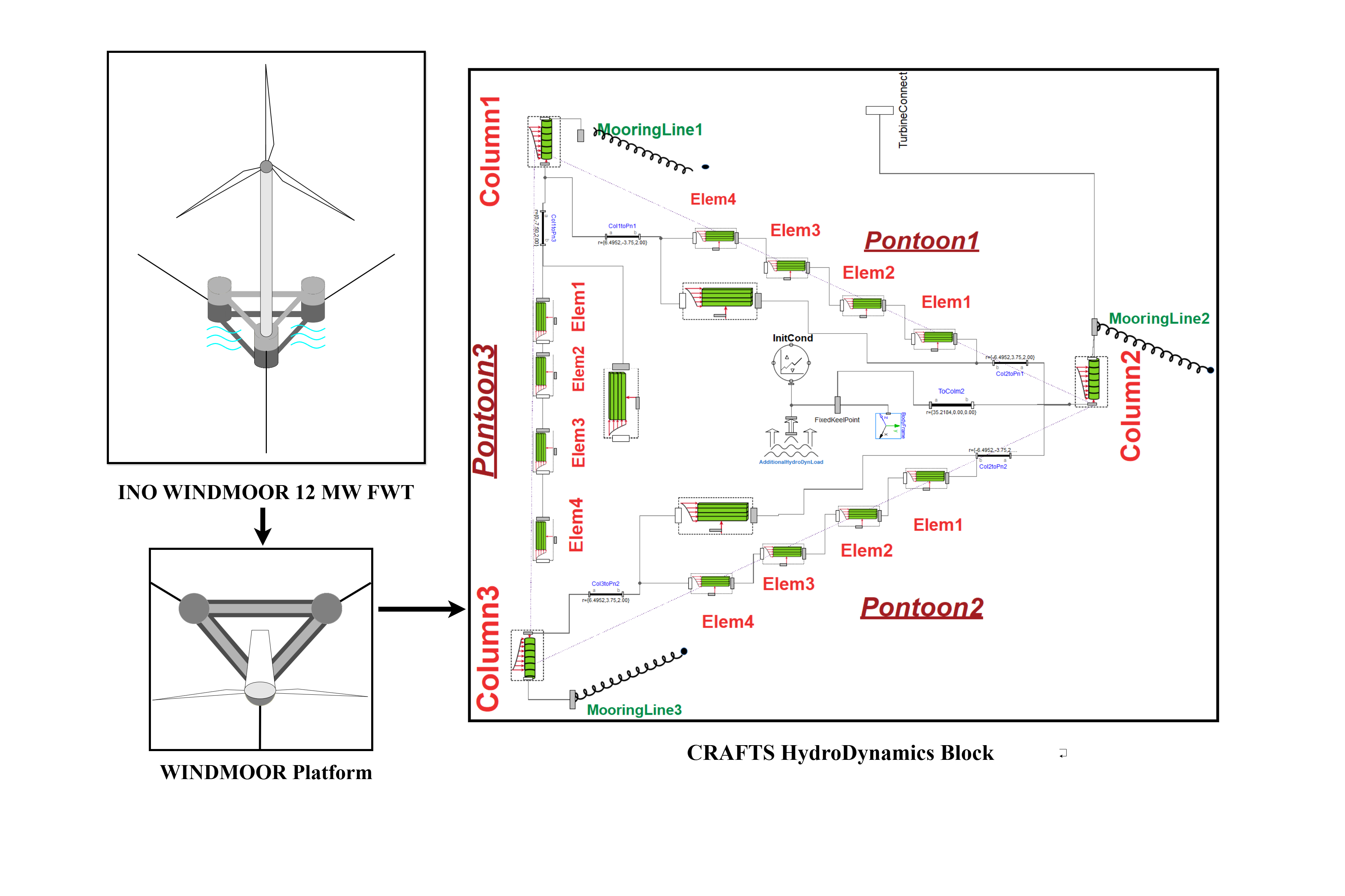}  
%\vspace{-0.05in}
\caption{Hydrodynamics model in CRAFTS} 
\label{fig:CRAFTS_hydro}
\end{center}
\end{figure*}
Modeling hydrodynamics is a critical challenge in predicting the dynamic behavior and load responses of FOWTs under varying sea-state conditions. Potential flow theory and Morison’s equation are widely used to evaluate hydrodynamic loads. Potential flow theory accounts for Froude-Krylov forces and diffraction effects for large rigid bodies but cannot capture drag forces due to its inviscid assumption. To overcome this limitation, a hybrid modeling approach combines potential-flow modeling with strip-theory drag force estimation.

In the Morison approach, accurately predicting hydrodynamic responses requires incorporating frequency-dependent drag coefficients, as the drag coefficient in oscillatory flows varies with the oscillation period. It was observed by \textbf{\cite{wang2022oc6}} that the axial-drag coefficients are more effective within specific frequency ranges, with lower frequencies requiring a lower drag coefficient. This frequency dependence arises from flow separation and vortex shedding at the heave plate corners, driven more by rapid wave-frequency oscillations than slower pitch resonance. Consequently, drag forces are more strongly influenced by wave-frequency velocity components, resulting in a higher effective drag coefficient at higher frequencies and a lower coefficient at lower frequencies. \textbf{\cite{wang2022oc6}} proposed a velocity filter approach using a first-order high-pass filter to incorporate the frequency dependency of the axial drag force, introducing an additional tunable parameter, $\alpha$,  which serves as a scaling factor for the axial drag coefficient. This method specifically improves the resonance frequency zone while leaving the wave frequency zone unaffected. 
% \textcolor{red}{Moreover, using a constant \(C_d\) value across the frequency band is difficult to accurately capture the response over both the resonance and wave-frequency ranges. @Sakif: double-check this stattement.} 
Additionally, \textbf{\cite{ishihara2019prediction}} proposed an augmented form of the Morison equation, incorporating correction factors for added mass and drag coefficients as functions of the oscillation period in an oscillating flow. The effectivenss of this approach was evaluated through free decay tests and water tank tests with regular and irregular waves.
% These limitations emphasize the need to introduce a new approach that can satisfy good predictions in both cases. 
On the other hand,\textbf{\cite{lemmer2018iterative}} proposed a method for implementing frequency-dependent drag coefficients in the frequency domain using linearized drag and an iterative solution. However, applying a suitable frequency-dependent drag coefficient in time-domain models remains a challenging task.

The main contribution of this study is a novel modeling approach that incorporates frequency-dependent hydrodynamic coefficients into Morison's equation using a set of frequency-dependent velocity filters. This method enables a more accurate prediction of the dynamic response of FOWTs under various sea states.
By addressing the challenge of incorporating frequency-dependent drag in time-domain models, this approach mitigates both under- and over-predictions across the frequency spectrum.  
The paper is structured as follows:
Section 2 outlines the hydrodynamic modeling framework, detailing the drag force calculation method with frequency-dependent drag coefficients. 
% The nonlinear hydrodynamic loads on the floater are modeled using the Morison equation, with wave kinematics derived from the linear Airy wave theory \textbf{\cite{chakrabarti1987hydrodynamics}}. 
% A comprehensive discussion on hydrodynamic modeling in CRAFTS can be found in \textbf{\cite{sarker2023causality}.
Section 3 presents the verification and validation of the frequency-dependent hydrodynamic model using experimental data from multiple load cases in the OC7 project, \cite{wang2025oc7definition,wang2025oc7}.  
Section 4 summarizes key findings, the effectiveness of frequency-dependent coefficients, and further works.

\section{Modeling and Simulation}

% \subsection{CRAFTS}
% CRAFTS is a multidomain, component-oriented cyberphysical simulation framework built on the Modelica language.It facilitates the acausal modeling of complex systems, such as floating offshore wind turbines (FOWTs), by integrating and simulating an assembly of fundamental physical components. It's drag and drop feature enables the simulation of various system configurations eliminating the need for extensive remodeling. CRAFTS consists of multiple interconnected components, including a floating platform, tower, rotor-nacelle assembly (RNA), and a mooring system along with aerodynamics,hydrodynamics and control modules. Structural dynamics module is modeled using a multi-body dynamics approach coupled with modal analysis theory, while the aerodynamics module employs blade element momentum theory. A brief description of the CRAFTS hydrodynamics module employed in this study is provided in the following subsections.

\subsection{Hydrodynamics Modeling}
The hydrodynamics model in this study computes the hydrodynamic loads on the floating platform using a relative form of the Morison equation. It accounts for key force components, including radiation-induced added mass, wave excitation, and non-linear viscous drag. 
%The hydrodynamic added mass and \textcolor{red}{drag coefficients are assumed to be independent of water depth (@No, drag coeffcient is depth dependent, we consider spalsh and submerge zone}), while diffraction effects are considered negligible. 
The hydrodynamic added mass is assumed to be independent of water depth, whereas the drag coefficient is considered to be depth-dependent due to the segmentation of the splash and submerged zones. Additionally, diffraction effects are considered negligible. The floating platform is discretized into multiple strip elements, with hydrodynamic forces on each element evaluated as follows:
\begin{equation}
    F = \frac{1}{4}\rho_{w} \pi D^2 \dot{u} + \frac{1}{4}\rho_{w} C_a \pi D^2 (\dot{u} - \dot{v}) + \frac{1}{2} \rho_{w} C_d D (u - v) |u - v|
\end{equation}

Where \( F \) denotes the force per unit length acting on each strip element, $\rho_{w}$ is the water density, and \( C_a \) and \( C_d \) correspond to the added mass and drag coefficients.
\( u\) and \( v \) represent the flow velocity and the velocity of the strip element, respectively.  \( D \) denotes the diameter of the strip element. This modeling approach allows users to easily drag and drop the basic components (e.g., cylinders, rectangular pontoons) to construct a floating platform (see Fig. \ref{fig:CRAFTS_hydro}). Each mooring line extends from the fairlead position on each outer column to a fixed anchor point. Unlike conventional catenary configurations, these lines are arranged horizontally and represented as massless linear springs \textbf{\cite{noboni2025modeling}}, providing an approximation of the stiffness characteristics of a catenary system. Interested readers can refer to our previous publication, \textbf{\cite{sarker2023causality}}, for further details. 

\begin{figure}[h]
\begin{center}
\includegraphics[width=8.4cm, trim={0cm 0cm 0cm 0cm},clip]{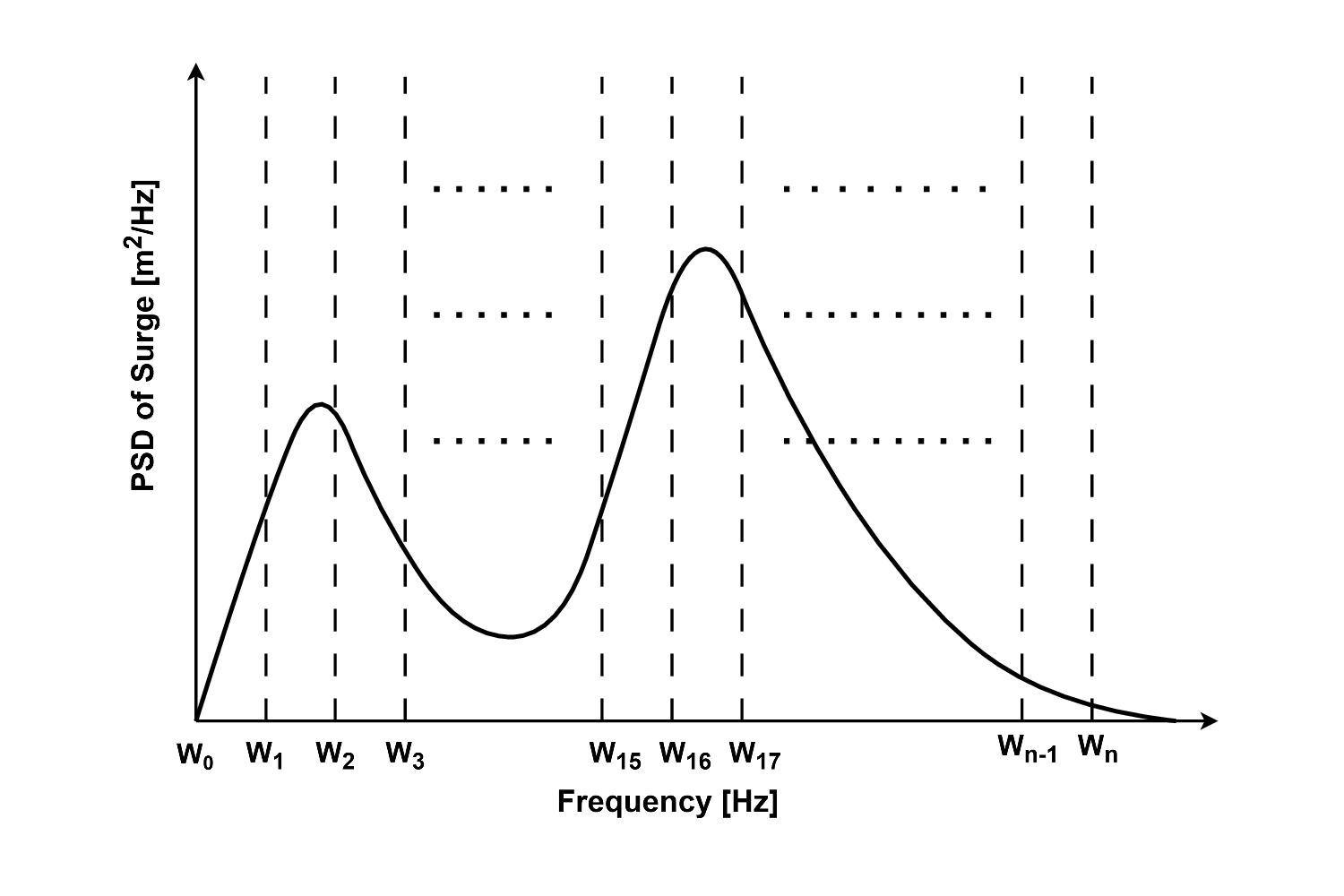}  
\caption{Surge PSD discretization} 
\label{fig:Surge_PSD}
\end{center}
\end{figure}

\subsection{Frequency-Dependent Drag Coefficients}

In this study, we modified the Morison equation to incorporate frequency-dependent hydrodynamic coefficients. To capture the frequency-dependent nature of the drag coefficients, we extended the original Morison equation, which uses the relative velocity term \(V_r = u - v\), by introducing a filtered velocity term. The modified drag force equation is expressed as:
\begin{equation}
F_d = \frac{1}{2} \cdot \rho_{w} \cdot A \cdot  \tilde{V}_r \cdot C_d \cdot |\tilde{V}_r|
\end{equation}
 \(F_d\) represents the drag force and  \(\tilde{V}_r\)  denotes the filtered relative velocity. To analyze a specific response, the Power Spectrum Density(PSD) of the motion is discretized into a multiple narrow frequency bands (see Fig. \ref{fig:Surge_PSD}).Each band corresponds to a distinct frequency range of the motion. A band-pass filter is applied to each band to isolate its respective frequency components \textbf{\cite{van1982analog}}. A second-order Butterworth band-pass filter is used to ensure a well-controlled frequency response. Butterworth filters provide a flatter passband and sharper roll-off than first-order filters,
making them more effective at preserving signal amplitude and waveform integrity within the
passband. The transfer function of the Butterworth filter is expressed as:
\begin{equation}
H(s) = \frac{\frac{\omega_c}{Q} \cdot s}{s^2 + \frac{\omega_c}{Q} s + \omega_c^2}
\end{equation}
where Q is the quality factor, which controls the sharpness of the filter, and \(w_c\) is the cutoff frequency. The cutoff frequencies of the Butterworth filter define the lower and upper limits of each frequency segment in the surge motion. The surge motion is decomposed into frequency segments, and the relative velocity \( V_r \) obtained from the hydrodynamics module is passed through band-pass filters to yield the filtered relative velocities \( \tilde{V}_r \) for each frequency range.
 For each segment, a specific drag coefficient 
\( C_d \) is assigned based on the corresponding frequency.
\begin{figure}[h]
\begin{center}
\includegraphics[width=8.4cm,height=11cm,trim={3.2cm 3.5cm 4.95cm 4cm},clip]{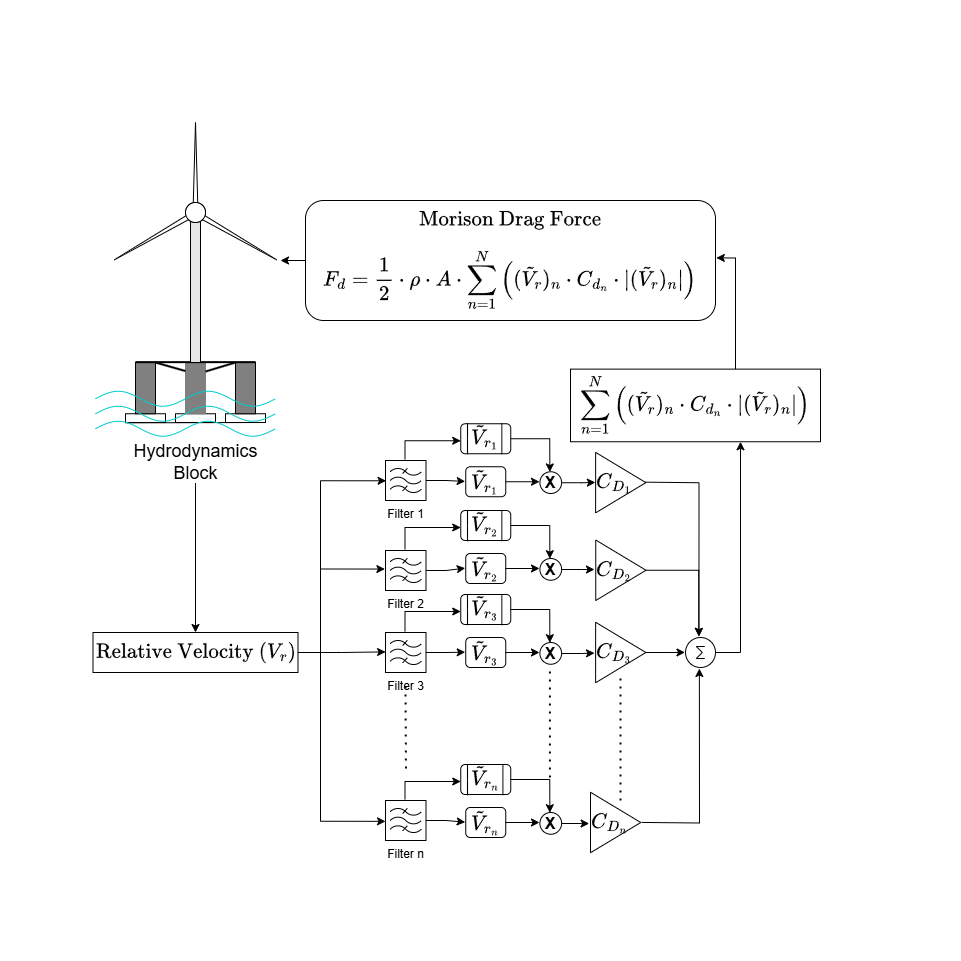}  
\caption{ Freq-Dependent Drag Coefficient Implementation} 
\label{fig:figure3}
\end{center}
\end{figure}
The force contribution from each frequency segment is computed by multiplying the filtered relative velocity 
\( \tilde{V}_r \) and its absolute value \(| \tilde{V}_r| \) with the corresponding drag coefficient \( C_d \), as shown in the following equation:
\begin{equation}
F_{d_n} = \frac{1}{2} \cdot \rho_{w} \cdot A \cdot \tilde{V}_{r_n} \cdot C_{d_n} \cdot |\tilde{V}_{r_n}|
\end{equation}
%Where \( F_{d_n} \) represents the drag force contribution from the \( n^{\text{th}} \) frequency segment. The total hydrodynamic drag force is obtained by summing the contributions from all frequency segments.
 Where \( F_{d_n} \) represents the drag force from the nth frequency segment, and the total drag force is,
\begin{equation}
F_{drag} = \sum_{n={1}}^{N} F_{d_n}
\end{equation}
Where \( N \) is the total number of frequency segments.

\section{SIMULATION AND VALIDATION RESULTS}
\subsection{Experiment Setup}
A series of load cases (LC) were evaluated for the INO WINDMOOR 12-MW wind turbine platform \textbf{\cite{thys2021experimental}} to validate the proposed frequency-dependent hydrodynamics model in this study. The floating platform comprises three cylindrical columns interconnected at the keel level by three rectangular pontoons arranged in a triangular configuration. The numerical results were compared with the experimental data obtained from SINTEF Ocean, conducted at a 1:40 scale, corresponding to a full-scale water depth of 150 m.
The prescribed load cases are categorized into two groups: (1) LC5.X – load cases for model tuning, and (2) LC6.X – load cases for blind validation. 
% \textcolor{red}{In all load cases, the global reference frame is defined with its origin at the intersection of the Mean Sea Water Level (MSWL) and the center of triangular domain}.  
\textcolor{black}{The validation process encompasses both time-domain and frequency-domain analyses to assess the platform's six degrees of freedom (surge, sway, heave, roll, pitch, and yaw)}.
\subsection{Simulation Setup}

%In this study, we analyze two distinct frequency ranges by segmenting the PSD plot into high- and low-frequency regions. The low-frequency range is associated with the resonance region, while the high-frequency range corresponds to the wave region. Different (\(C_d\)) values are assigned: one for the low-frequency range and one for the high-frequency range. To implement this approach, two Butterworth filters are applied: a fifth-order low-pass Butterworth filter for the resonance frequency zone, and a first-order high-pass Butterworth filter for the wave frequency zone, both with a cutoff frequency of 0.05. The platform's columns are divided into two distinct zones - splash and submerged - with separate transverse drag coefficients (\(C_d\)) assigned to each. In total, six tunable \(C_d\) values are evaluated: four for the columns and two for the pontoon in the surge direction. The main objective of this study is to demonstrate how frequency-dependent drag coefficients improve the accuracy of the hydrodynamics model. The proposed frequency-dependent drag model is applied only in the surge direction, while for the heave motion, a single \(C_d\) value is applied across the entire frequency range, with two vertical drag coefficients (\(C_d\)) assigned to each column and pontoon.

In this study, we analyze two distinct frequency ranges by segmenting the PSD plot into high- and low-frequency regions. The low-frequency range is associated with the resonance region, while the high-frequency range corresponds to the wave region. Different (\(C_d\)) values are assigned: one for the low-frequency range and one for the high-frequency range. To implement this approach, two Butterworth filters are applied: a fifth-order low-pass Butterworth filter for the resonance frequency zone and a first-order high-pass Butterworth filter for the wave frequency zone, both with a cutoff frequency of 0.05. The fifth-order filter is employed in the low-frequency zone to reduce the effects of the wave frequency zone. In contrast, the resonance frequency zone has minimal impact in the wave frequency region, making a first-order filter adequate for this application. The platform's columns are divided into two distinct zones - splash and submerged - with separate transverse drag coefficients (\(C_d\)) assigned to each. In total, six tunable \(C_d\) values are evaluated: four for the columns and two for the pontoon in the surge direction. The main objective of this study is to demonstrate how frequency-dependent drag coefficients improve the accuracy of the hydrodynamics model. The proposed frequency-dependent drag model is applied only in the surge direction, while for the heave motion, a single \(C_d\) value is applied across the entire frequency range, with two vertical drag coefficients (\(C_d\)) assigned to each column and pontoon.
\subsection{Load Case 5.X for Tuning Coefficients}
\begin{table}[h]
    \centering
    \caption{Load cases 5.X for tuning coefficients}
    
    \begin{tabular}{ccccc}
        \toprule
        Load Case & $H_s$ [m] & $T_p$ [s] & $\gamma$ [-] & $H_S/D$ \\
        \midrule
        5.0           & 2.00 & 7.0 & -    & 0.13  \\
        5.1           & 3.74 & 7.0 & 4.90 & 0.25  \\
        5.2           & 6.19 & 9.0 & 4.90 & 0.41  \\
        5.3           & 11.0 & 12  & 4.90 & 0.73  \\
        5.4           & 3.74 & 12  & 1.00 & 0.25  \\
        
        \bottomrule
    \end{tabular}
\end{table}
\begin{figure}[h]
\begin{center}
\includegraphics[width=8.4cm,trim={0.6cm 1.4cm 0cm 0.5cm},clip]{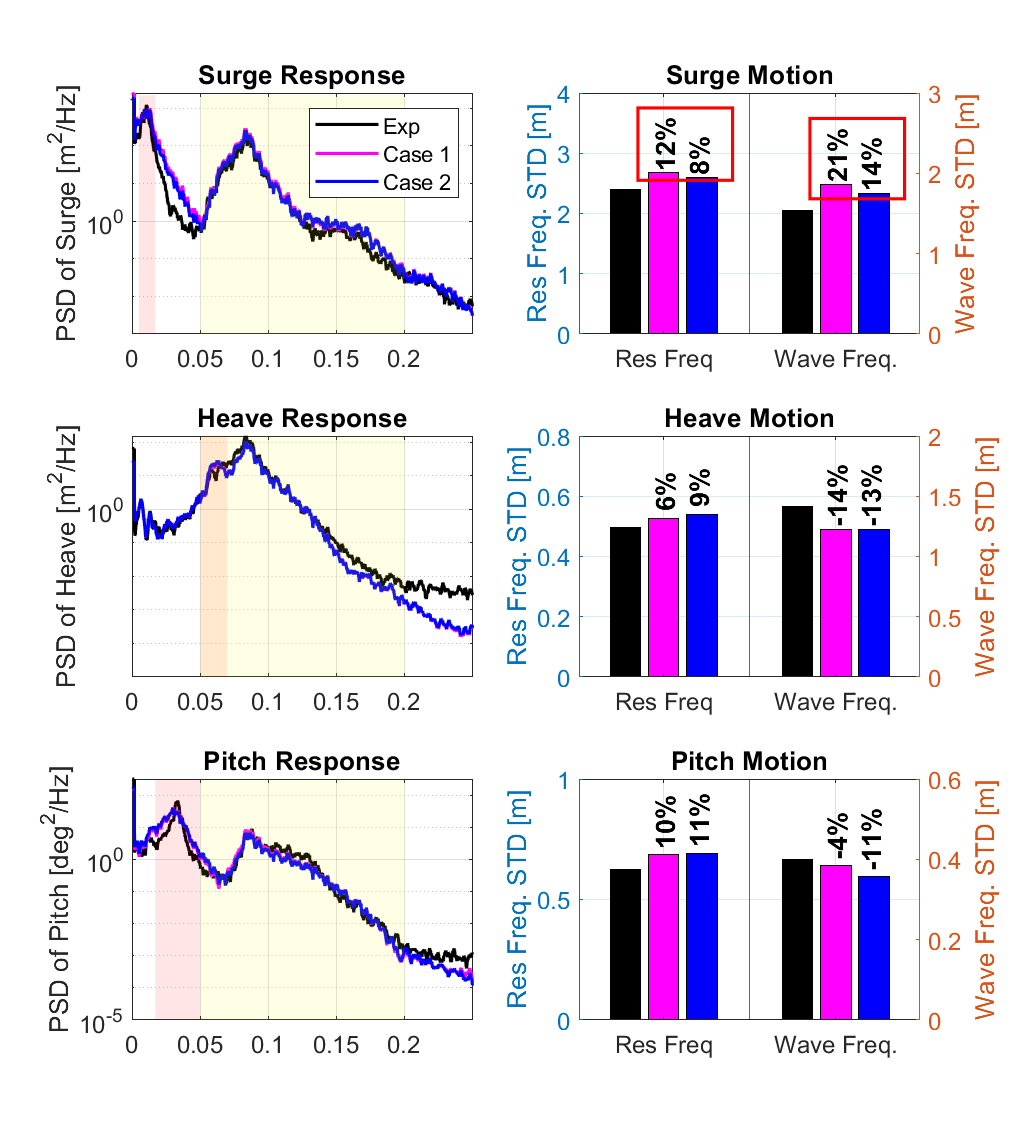}  
\caption{PSD of surge, heave, pitch in LC 5.3 (Case 1: constant drag coefficient, Case 2: frequency-dependent drag coefficients)} 
\label{fig:LC53}
\end{center}
\end{figure}

In Load Case 5.X, the responses of the platform, including surge, heave, and pitch motions, are analyzed under irregular wave conditions characterized by the JONSWAP wave spectrum. This study examines a series of load cases (LC 5.X) from Table 1, covering a wide range of significant wave heights (Hs) and spectral peak periods (Tp). All load cases are subjected to a 0-degree wave heading where the waves approach directly towards the platform.\begin{figure}[h]
\begin{center}
\includegraphics[width=8.4cm, trim={0.8cm 1.4cm 0cm 0.5cm},clip]{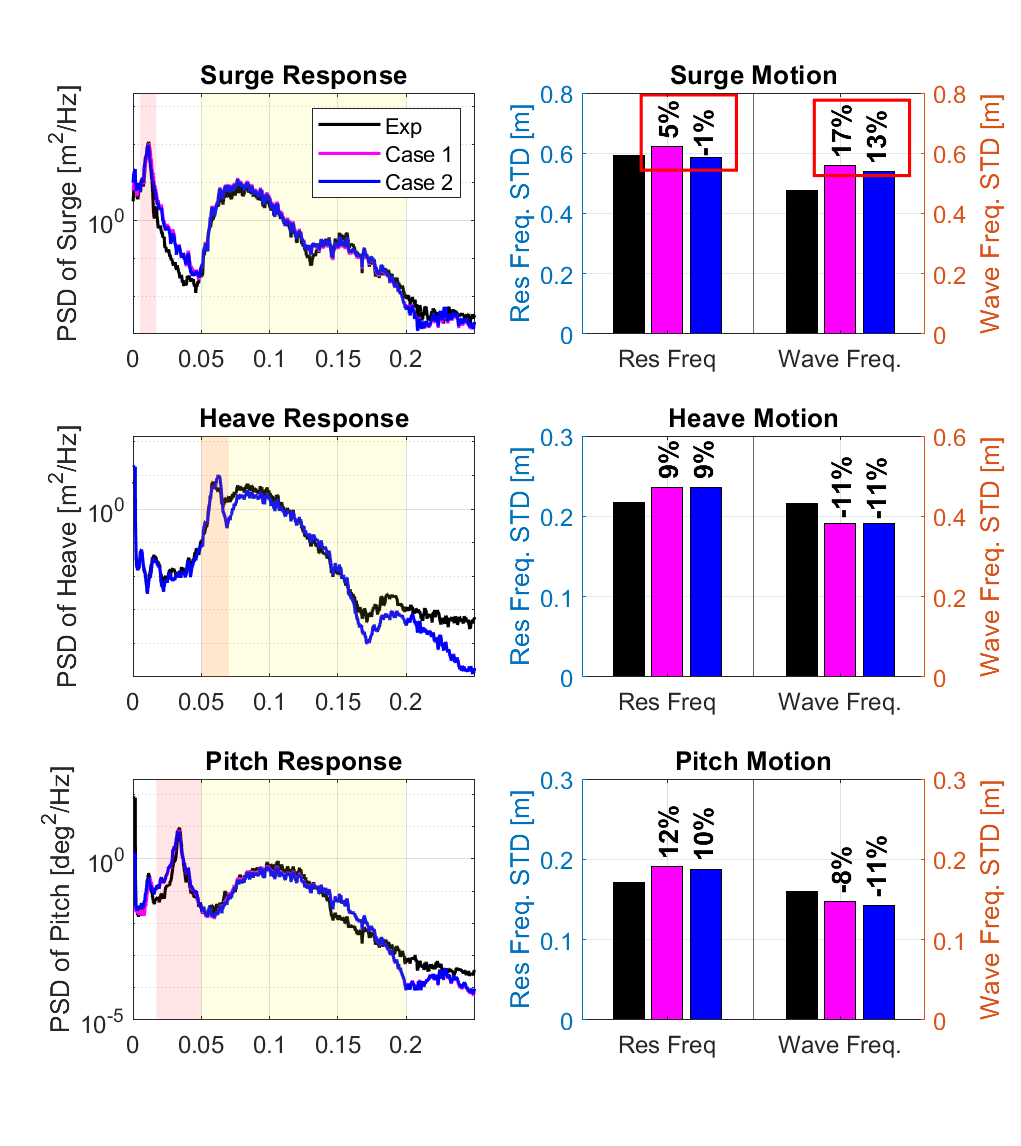}  
\caption{PSD of surge, heave, pitch in LC 5.4 (Case 1: constant drag coefficient, Case 2: frequency-dependent drag coefficients)} 
\label{fig:LC54}
\end{center}
\end{figure}
 Numerical simulations were conducted for 12000 seconds, and the resulting data were analyzed using power spectral density (PSD) to assess system's dynamic behavior. This investigation evaluates the impact of frequency-dependent drag modeling on the platform’s responses across both low- and high-frequency ranges.

%%
%\begin{figure}[h]
%\begin{center}
%\includegraphics[width=8.4cm, trim={0.7cm 1.2cm 0cm 0.8cm},clip]{LC50.eps}  
%\caption{PSD of surge, heave, pitch in LC 5.0} 
%\label{fig:LC50}
%\end{center}
%\end{figure}

%%
%\begin{figure}[h]
%\begin{center}
%\includegraphics[width=8.4cm, trim={0.7cm 0.8cm 0cm 0.8cm},clip]{LC51.eps}  
%\caption{PSD of surge, heave, pitch in LC 5.1} 
%\label{fig:LC51}
%\end{center}
%\end{figure}

%%
%\begin{figure}[h]
%\begin{center}
%\includegraphics[width=8.4cm, trim={0.7cm 0.8cm 0cm 1cm},clip]{LC52.eps}  
%\caption{PSD of surge, heave, pitch in LC 5.2} 
%\label{fig:LC52}
%\end{center}
%\end{figure}
%%

%%
\begin{table}[h]
    \centering
    \caption{LC 5.X Surge motion STD Error Summary}
    \label{pass_rate_summary1}
    \begin{tabular}{c c c c c }
        \toprule
        & \multicolumn{2}{c}{Case 1} & \multicolumn{2}{c}{Case 2}  \\
        \cmidrule(lr){2-3} \cmidrule(lr){4-5}
        Load Case & Res. & Wave & Res. & Wave  \\
        \midrule
        5.0 & 2\% & -1\% & 0\% & 0\%  \\
        5.1 & -6\% & -3\% & 0\% & 0\%  \\
        5.2 & 7\% & 9\% & 1\% & 8\%  \\
        5.3 & 12\% & 21\% & 8\% & 14\%  \\
        5.4 & 5\% & 17\% & -1\% & 13\%  \\

        \bottomrule
    \end{tabular}
\end{table}

In Load Case 5.X, two simulation cases were compared against experimental data: Case 1, which does not incorporate frequency-dependent drag coefficients, and Case 2, which includes frequency-dependent drag modeling. Due to page limitations, only the plots for  LC 5.3 and LC 5.4 are presented. Note that the right-hand side of the plot shows the standard deviation (STD) of the motion, calculated as the square root of the PSD integral over the resonance frequencies and wave frequencies, highlighted by the shaded areas in the left-hand side plot. The results for the surge motion in LC 5.X for both cases are summarized in Table \ref{pass_rate_summary1}. In LC 5.0 to 5.1, Case 2 demonstrates good agreement with experimental results for surge in both low- and high-frequency region with an STD error of 0\% in both cases (see table \ref{pass_rate_summary1}). 
% \textcolor{red}{@Sakif: provide a definition of PSD sum and the intergral interval}. 
In LC 5.2 (see table \ref{pass_rate_summary1}), the error in resonance frequency reduced from 7\% to 1\%, while in the wave frequency region, the error reduction is only 1\%. Furthermore, in LC 5.3 (Fig. \ref{fig:LC53}) and 5.4 (Fig. \ref{fig:LC54}), Case 1 overestimates the surge response in the high-frequency range, leading to deviations beyond the acceptable range. An acceptance threshold of ±15\% in the square roots of the PSD sum is applied to all metrics and load cases.
% \textcolor{red}{@Sakif: give a definition of the acceptable range.} 
By incorporating frequency-dependent drag modeling in Case 2, the results show better agreement with experimental data in both the resonance and wave frequency ranges for the surge motion, demonstrating the effectiveness of the proposed modeling approach.

% \subsection{Load Case 6.X}
% \begin{table}[h]
%     \centering
%     \caption{LC 6.X for blind validation}
%     \label{LC6X}
%     \begin{tabular}{ccccc}
%         \toprule
%         Load Case & $H_s$ [m] & $T_p$ [s] & $\gamma$ [-] & $H_S/D$ \\
%         \midrule
%         6.1           & 3.74 & 9.0 & 1.49    & 0.25  \\
%         6.2           & 6.19 & 9.0 & 4.90 & 0.41  \\
%         6.3           & 11.0 & 12 & 4.90 & 0.73  \\
%         6.4           & 6.19 & 12  & 1.23 & 0.41  \\

%         6.7           & 6.19 & 12  & 1.23 & 0.41  \\
        
%         \bottomrule
%     \end{tabular}
% \end{table}

\subsection{Load Case 6.1 for Blind Validation}
\begin{table}[h]
    \centering
    \caption{LC 6.1 for blind validation}
    \label{LC6X}
    \begin{tabular}{ccccc}
        \toprule
        Load Case & $H_s$ [m] & $T_p$ [s] & $\gamma$ [-] & $H_S/D$ \\
        \midrule
        6.1           & 3.74 & 9.0 & 1.49    & 0.25  \\

        \bottomrule
    \end{tabular}
\end{table}

% \subsection{Load Case 6.X}
% \begin{table}[h]
%     \centering
%     \caption{LC 6.X for blind validation}
%     \label{LC6X}
%     \begin{tabular}{ccccc}
%         \toprule
%         Load Case & $H_s$ [m] & $T_p$ [s] & $\gamma$ [-] & $H_S/D$ \\
%         \midrule
%         6.1           & 3.74 & 9.0 & 1.49    & 0.25  \\
%         6.2           & 6.19 & 9.0 & 4.90 & 0.41  \\
%         6.3           & 11.0 & 12 & 4.90 & 0.73  \\
%         6.4           & 6.19 & 12  & 1.23 & 0.41  \\
%         6.5           & 14.97 & 14  & 4.9 & 0.99  \\
%         6.7           & 6.19 & 12  & 1.23 & 0.41  \\
        
%         \bottomrule
%     \end{tabular}
% \end{table}

% With the exception of LC 6.2 and LC 6.3, all LC 6.X cases correspond to distinct sea states, characterized by variations in significant wave height ($H_s$) and spectral peak period ($T_p$), compared to those in LC 5.X. LC 6.2 and LC 6.3 share the same sea states as LC 5.2 and LC 5.3 but feature different wave realizations to assess whether models calibrated for a given sea state can accurately predict system responses under varying wave conditions. 
% Notably, LC 6.5 represents a an extreme sea state beyond the range considered in LC 5.X, necessitating the extrapolation of model coefficients. 
% Additionally, LC 6.7 introduces a constant tower-top thrust force, inducing a mean pitch offset. This case aims to examine the impact of mean pitch offset on the platform's hydrodynamic responses. Due to page limitations, we only provide simulation plots for LC 6.1. 

%%
\begin{figure}[h]
\begin{center}
\includegraphics[width=8.4cm, trim={1.5cm 0.8cm 1.5cm 0cm},clip]{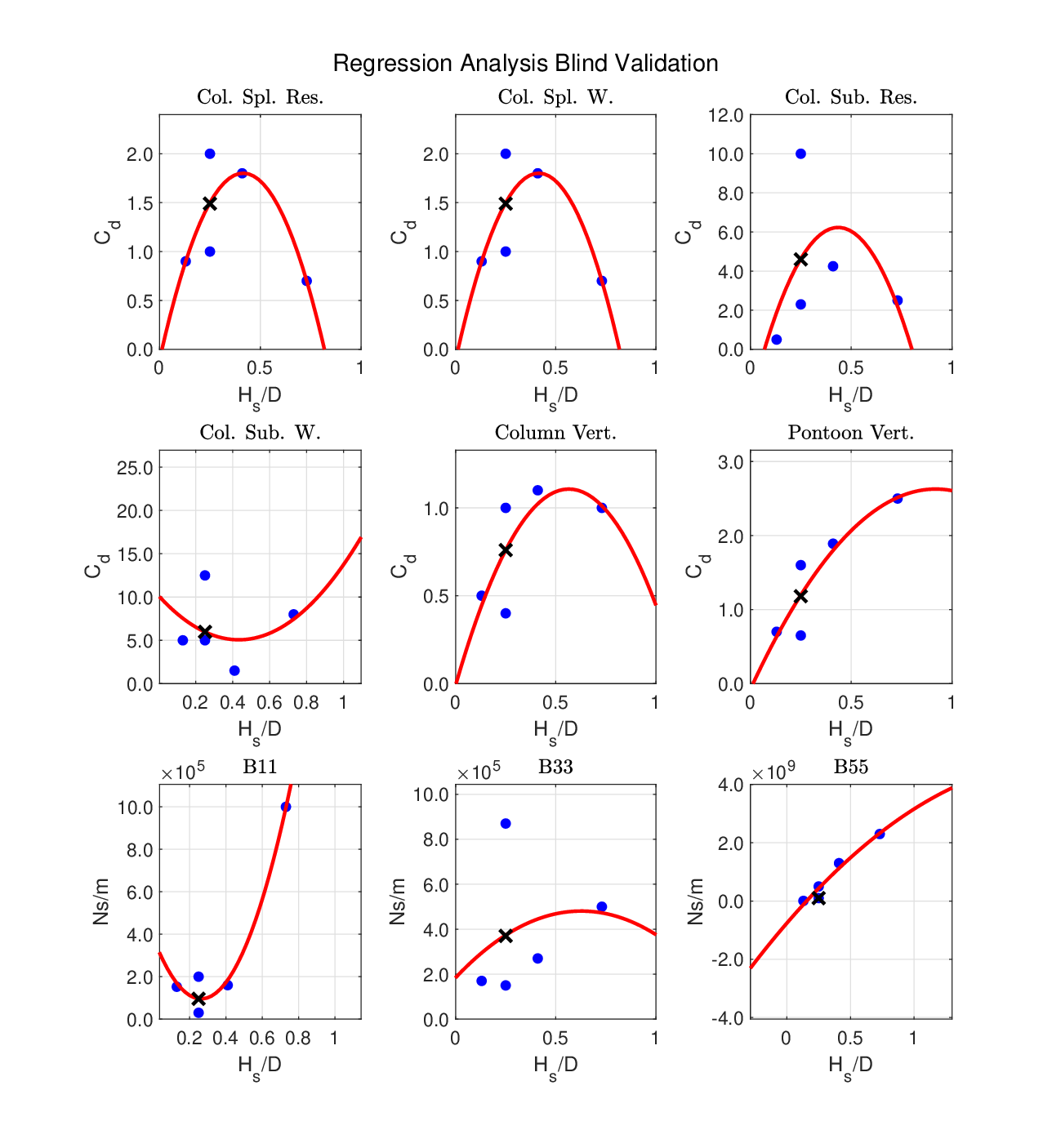}  
\vspace{-.05in}
\caption{Regression models (blue: LC 5.X, black: LC 6.1)} 
\label{fig:reg_model}
\end{center}
\end{figure}

\begin{figure}[h]
\begin{center}
\includegraphics[width=8.4cm, height=9 cm,trim={0.7cm 1.5cm 0cm 1cm},clip]{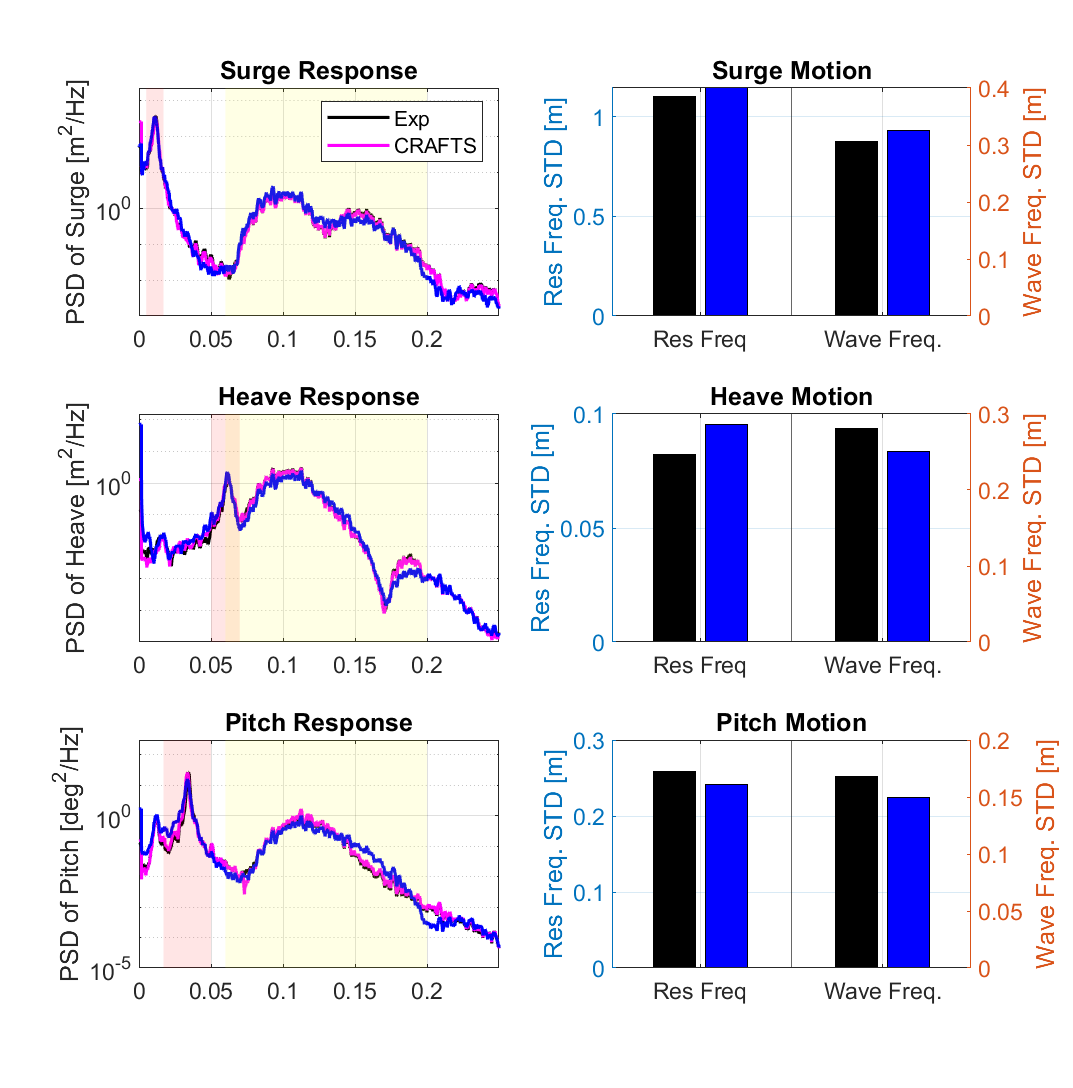}  
\caption{PSD of surge, heave, pitch in LC 6.1} 
\label{fig:LC61}
\end{center}
\end{figure}

Table \ref{LC6X}  presents the load case 6.1 designated for blind validation.                 
For blind validation, a regression analysis was performed by plotting \( C_d \) against  the normalized significant wave height, \( H_s/D \),  which is approximately proportional to the Keulegan–Carpenter (KC) number, using data from the load case 5.X. A quadratic fit was applied to derive the regression model (see Fig. \ref{fig:reg_model}). For LC 6.1, where \( H_s/D \) is 0.25, the corresponding \( C_d \) values were obtained through interpolation between Load Cases 5.1 and 5.4 using the regression model. Fig. \ref{fig:LC61} shows an error of 6\%  at the resonance frequency and 7\% at the wave frequency, which closely aligns with the experimental data. Table \ref{pass_rate_summary} presents the modeling error summary for LC 5.X and LC 6.1. This further validates the effectiveness of the proposed approach in predicting system responses under varying sea states.  

\begin{table}[h]
    \centering
    \caption{STD Error Summary}
    \label{pass_rate_summary}
    \begin{tabular}{c c c c c c c}
        \toprule
        & \multicolumn{2}{c}{Surge} & \multicolumn{2}{c}{Heave} & \multicolumn{2}{c}{Pitch} \\
        \cmidrule(lr){2-3} \cmidrule(lr){4-5} \cmidrule(lr){6-7}
        Load Case & Res. & Wave & Res. & Wave & Res. & Wave \\
        \midrule
        5.0 & 0\% & 0\% & 5\% & 6\% & -1\% & 6\% \\
        5.1 & 0\% & 0\% & 2\% & 7\% &  1\% & 15\% \\
        5.2 & 1\% & 8\% & 2\% & -5\% &  5\% & -14\% \\
        5.3 & 8\% & 14\% & 9\% & -13\% &  11\% & -11\% \\
        5.4 & -1\% & 13\% & 9\% & -11\% &  10\% & -11\% \\
        6.1 & 5\% & 7\% & 0\% & -8\% &  -11\% & -11\% \\

        \bottomrule
    \end{tabular}
\end{table}
\section{Conclusion}

A frequency-dependent drag model has been successfully developed and validated against the experimental data for the INO WINDMOOR 12-MW wind turbine platform.
The results show significant improvements in the prediction of surge resonance and wave frequency responses. Although a slight overprediction persists in the surge wave response for LC 5.3 and 5.4, this can be mitigated by employing multiple velocity filters across narrow frequency bands. In blind validation, the model exhibits only a slight 5\% and 6\% overprediction in the resonance and wave frequency responses, respectively, indicating strong agreement with the experimental data. These findings confirm the effectiveness of the proposed modeling approach in enhancing the hydrodynamic response predictions. The pass rates for LC 5.X improve from 85\% (without frequency-dependent hydrodynamic coefficients) to 100\% when using the proposed approach, confirming its effectiveness in improving hydrodynamic response predictions. For LC 6.1, blind validation using a regression-based estimation of hydrodynamic coefficients across various sea states also yielded accurate predictions of FOWT responses. 
While heave and pitch responses remain satisfactory without incorporating frequency-dependent drag coefficients, implementing this model in the vertical direction is expected to further improve heave response, as well as pitch response, given that pitch is influenced by both surge and heave motions. Future work will explore using a physics-informed neural network to predict the \(C_d\) values for each frequency band, thereby eliminating the need for manual tuning procedures.    

\begin{ack}
The experimental data from the WINDMOOR campaign, funded by the Research Council of Norway (Grant No. 321954) and industry partners and conducted at the SINTEF Ocean basin, have been provided by SINTEF Ocean for use within the OC7 project. The authors also thank National Renewable Energy Laboratory (NREL) for the valuable opportunity to participate in the OC7 project and validate their hydrodynamic model.
\end{ack}

\bibliography{ifacconf}             % bib file to produce the bibliography

\begin{thebibliography}{15}
\providecommand{\natexlab}[1]{#1}
\providecommand{\url}[1]{\texttt{#1}}
\providecommand{\urlprefix}{URL }
\expandafter\ifx\csname urlstyle\endcsname\relax
  \providecommand{\doi}[1]{doi:\discretionary{}{}{}#1}\else
  \providecommand{\doi}{doi:\discretionary{}{}{}\begingroup \urlstyle{rm}\Url}\fi

\bibitem[{Garcia-Sanz(2019)}]{garcia2019control}
Garcia-Sanz, M. (2019).
\newblock Control co-design: an engineering game changer.
\newblock \emph{Advanced Control for Applications: Engineering and Industrial Systems}, 1(1), e18.

\bibitem[{Ishihara and Zhang(2019)}]{ishihara2019prediction}
Ishihara, T. and Zhang, S. (2019).
\newblock Prediction of dynamic response of semi-submersible floating offshore wind turbine using augmented morison's equation with frequency dependent hydrodynamic coefficients.
\newblock \emph{Renewable energy}, 131, 1186--1207.

\bibitem[{Jonkman et~al.(2022)Jonkman, Mudafort, Platt, Branlard, Sprague, Jonkman, Hayman, Vijayakumar, Buhl, Ross et~al.}]{jonkman2022openfast}
Jonkman, B., Mudafort, R., Platt, A., Branlard, E., Sprague, M., Jonkman, J., Hayman, G., Vijayakumar, G., Buhl, M., Ross, H., et~al. (2022).
\newblock Openfast/openfast: Openfast v3. 1.0.
\newblock \emph{Zenodo [code]}, 10.

\bibitem[{Jung et~al.(2024)Jung, Sander, and Schindler}]{jung2024future}
Jung, C., Sander, L., and Schindler, D. (2024).
\newblock Future global offshore wind energy under climate change and advanced wind turbine technology.
\newblock \emph{Energy Conversion and Management}, 321, 119075.

\bibitem[{Lemmer et~al.(2018)Lemmer, Yu, and Cheng}]{lemmer2018iterative}
Lemmer, F., Yu, W., and Cheng, P.W. (2018).
\newblock Iterative frequency-domain response of floating offshore wind turbines with parametric drag.
\newblock \emph{Journal of Marine Science and Engineering}, 6(4), 118.

\bibitem[{Mohsin et~al.(2025)Mohsin, Odeh, Ngo, and Das}]{mohsin2025dynamically}
Mohsin, K., Odeh, M., Ngo, T., and Das, T. (2025).
\newblock Dynamically allocated individual pitch control for fatigue load reduction in wind turbines.
\newblock \emph{Control Engineering Practice}, 161, 106357.

\bibitem[{Noboni et~al.(2025)Noboni, McConnell, and Das}]{noboni2025modeling}
Noboni, T., McConnell, J., and Das, T. (2025).
\newblock Modeling tethered multirotor autogyro with altitude control via differential rotor braking.
\newblock \emph{Journal of Guidance, Control, and Dynamics}, 1--14.

\bibitem[{Odeh et~al.(2023)Odeh, Mohsin, Ngo, Zalkind, Jonkman, Wright, Robertson, and Das}]{odeh2023development}
Odeh, M., Mohsin, K., Ngo, T., Zalkind, D., Jonkman, J., Wright, A., Robertson, A., and Das, T. (2023).
\newblock Development of a wind turbine model and simulation platform using an acausal approach: Multiphysics modeling, validation, and control.
\newblock \emph{Wind Energy}, 26(9), 985--1011.

\bibitem[{Sarker et~al.(2023)Sarker, Hasan, Ngo, and Das}]{sarker2023causality}
Sarker, D., Hasan, T., Ngo, T., and Das, T. (2023).
\newblock Causality-free modeling of a floating wind turbine semisubmersible platform with validation results.
\newblock \emph{IFAC-PapersOnLine}, 56(3), 559--564.

\bibitem[{Sarker et~al.(2024)Sarker, Tran, Mohsin, Odeh, Ngo, and Das}]{sarker2024modeling}
Sarker, D., Tran, D., Mohsin, K., Odeh, M., Ngo, T., and Das, T. (2024).
\newblock Modeling, validation, and control of the iea-15mw reference wind turbine and volturnus-s platform.
\newblock \emph{IFAC-PapersOnLine}, 58(28), 1--6.

\bibitem[{Thys et~al.(2021)Thys, Souza, Sauder, Fonseca, Berthelsen, Engebretsen, and Haslum}]{thys2021experimental}
Thys, M., Souza, C., Sauder, T., Fonseca, N., Berthelsen, P.A., Engebretsen, E., and Haslum, H. (2021).
\newblock Experimental investigation of the coupling between aero-and hydrodynamical loads on a 12 mw semi-submersible floating wind turbine.
\newblock In \emph{International Conference on Offshore Mechanics and Arctic Engineering}, volume 85192, V009T09A030. American Society of Mechanical Engineers.

\bibitem[{Van~Valkenburg(1982)}]{van1982analog}
Van~Valkenburg, M.E. (1982).
\newblock Analog filter design.
\newblock \emph{(No Title)}.

\bibitem[{Wang et~al.(2025{\natexlab{a}})Wang, Robertson, Jonkman, Berthelsen, and Thys}]{wang2025oc7definition}
Wang, L., Robertson, A., Jonkman, J., Berthelsen, P.A., and Thys, M. (2025{\natexlab{a}}).
\newblock {OC7 Phase I Definition Document}.
\newblock Technical report, National Renewable Energy Laboratory, Golden, CO, USA.

\bibitem[{Wang et~al.(2025{\natexlab{b}})Wang, Robertson, Jonkman, Liao, Berthelsen, Abdelmoteleb, Rohrer, Ramachandran Nair~Rajasree, Bachynski-Polić, Clement et~al.}]{wang2025oc7}
Wang, L., Robertson, A., Jonkman, J., Liao, Y., Berthelsen, P.A., Abdelmoteleb, S.E., Rohrer, P., Ramachandran Nair~Rajasree, V., Bachynski-Polić, E., Clement, C., et~al. (2025{\natexlab{b}}).
\newblock {OC7 Phase I: Toward practical sea-state-dependent modeling of hydrodynamic viscous drag and damping}.
\newblock \emph{Ocean Engineering}.

\bibitem[{Wang et~al.(2022)Wang, Robertson, Jonkman, and Yu}]{wang2022oc6}
Wang, L., Robertson, A., Jonkman, J., and Yu, Y.H. (2022).
\newblock Oc6 phase i: Improvements to the openfast predictions of nonlinear, low-frequency responses of a floating offshore wind turbine platform.
\newblock \emph{Renewable Energy}, 187, 282--301.

\end{thebibliography}
                                                     % with bibtex (preferred)

    % in the appendices.
\end{document}